\documentclass[a4paper,11pt]{article}
\usepackage{calc}
\usepackage[text={17.5cm,25.5cm},centering]{geometry}

\usepackage{epsfig}
\usepackage{amsmath}
\usepackage{verbatim}
\usepackage{amssymb}
\usepackage{bbm}
\usepackage{graphicx}
\usepackage{textcomp}
\usepackage{listings}
\usepackage{here}
\numberwithin{equation}{section}
\usepackage[english]{babel}        
\usepackage{epstopdf}

\usepackage[applemac]{inputenc}
\usepackage{units}
\usepackage {subfig}
\usepackage{pdfpages}
\usepackage{booktabs}

\usepackage[font=small,labelfont=bf]{caption}
\usepackage[calcwidth]{titlesec}

\usepackage[noblocks]{authblk}

\begin{document}

\title{Spatial features of synaptic adaptation affecting learning performance}

\author[1,*]{Damian L. Berger}
\author[2,3]{Lucilla de Arcangelis}
\author[1]{Hans J. Herrmann}
\affil[1]{ETH Zürich, Computational Physics for Engineering Materials, IfB, Wolfgang-Pauli-Strasse 27, 8093 Zürich, Switzerland}
\affil[2]{Department of Industrial and Information Engineering, University of Campania "Luigi  Vanvitelli", 81031 Aversa (CE), Italy}
\affil[3]{INFN sez. Naples, Gr. Coll. Salerno (Italy)}

\affil[*]{bergerda@ethz.ch}


\maketitle

\begin{abstract}
Recent studies have proposed that the diffusion of messenger molecules, such as monoamines, can mediate the plastic adaptation of synapses in supervised learning of neural networks. Based on these findings we developed a model for neural learning, where the signal for plastic adaptation is assumed to propagate through the extracellular space. We investigate the conditions allowing learning of Boolean rules in a neural network. Even fully excitatory networks show very good learning performances. Moreover, the investigation of the plastic adaptation features optimizing the performance suggests that learning is very sensitive to the extent of the plastic adaptation and the spatial range of synaptic connections.
\end{abstract}

%
%

\section*{Introduction}

The brain is a large neural network, where sensory information is processed, possibly triggering a specific response. If this response turns out to generate a result, that is different from the expected one, error signals cause changes in the synaptic weights attempting to minimize further mismatches \cite{schultz2000neuronal,keller2009neural,keller2012sensorimotor}. Models of neural network attempting to contribute to the understanding of learning in the brain usually look at a task where for a specific set of inputs the network has to adapt to generate a desired result. Error back propagation \cite{Backprop}, one of the most popular algorithms proposed, can however not serve as a candidate for learning in a biological sense \cite{Crick}. A teaching signal which transmits information antidromically with memory of all neurons it has passed before, seems unlikely to exist in the brain. 

Different studies have proposed a variety of models for a neural network able to learn without passing error gradients antidromically. It has been suggested that errors could propagate backwards through a second network \cite{kolen1994backpropagation,stork1989backpropagation}, where forward and backward connections have to be chosen symmetric. Another interesting approach is to calculate error gradients by using an alternation of two activity phases, where one is determined by a teacher \cite{o1996biologically}. A similar performance, as the back propagation algorithm, can also be achieved by using exact error gradients only for the last synapses and the synapses in the hidden layers are modified only in the same direction as with back propagation \cite{lillicrap2016random}. There exist also interesting models, called liquid state machines or echo state machines, where the weights of the main network remains unchanged and the inputs are connected to a large recurrent neural network, thereby mapping the input to a higher dimensional space \cite{maass2002real,jaeger2001echo}. These models are mainly used for spatiotemporal problems like speech recognition, but their capability for recognizing handwritten digits has also been demonstrated \cite{jalalvand2015real}. 

The variety of model reduces if we insist on the assumption that synaptic changes have to be triggered by a specific error signal. Since there is no evidence that a neural network can transfer feedback information antidromically, such a signal should propagate through the extracellular space. A feedback signal to change the strength of  synapses in a larger region could for instance be delivered by monoamine releasing neurons. It is known that these neurons release their transmitters deep into the extracellular space \cite{VT}. In particular for dopamine, it has been verified, that its release can mediate plasticity \cite{D1D5, Dopamine}. In several recently proposed models these ideas are implemented by the assumption that spike-timing-dependent plasticity (STDP) and dopamine induced plasticity are directly coupled \cite{STDPdopamine, Florian,aswolinskiy2015rm}. These models assume that dopamine serves as a rewarding signal. In an other model it has been proposed \cite{Seung} that the transmitted feedback signal changes the vesicle release probability of previously activated synapses. 

Opposite to rewarding reinforcement it has been argued that negative feedback signals, which change synapses only if mistakes occurred, are more biologically plausible and preserve the adaptability of the system \cite{BakChialvo, chialvo1999learning, Bosman}. Based on such ideas, de Arcangelis et al.\cite{LucillaCrit, LucillaOptimal} proposed a model of spiking integrate-and-fire neurons which is able to learn logical binary functions in a neural network. For wrong responses by the neural network, activated synapses are modified proportionally to the inverse number of synaptic connections on the signal path to the output neuron. This method implies that the feedback signal propagates backwards on the synaptic connections and decreases in strength depending on the number of neurons it has passed before. In this article we extend these ideas by a model, where the strength of the feedback signal does not depend on a network distance, but on the euclidean distance. By this model we study the importance of the localization of a teaching signal, i.e. if a localized learning signal can represent advantages over one acting widely in space. This is the case, for instance, of dopamine whose effect covers a finite range of tens to thousands of synapses \cite{VT}.

\section*{Network model}

\subsection*{Network topology}
\begin{figure}[htbp]
  \centering
    \includegraphics[width=0.5\textwidth]{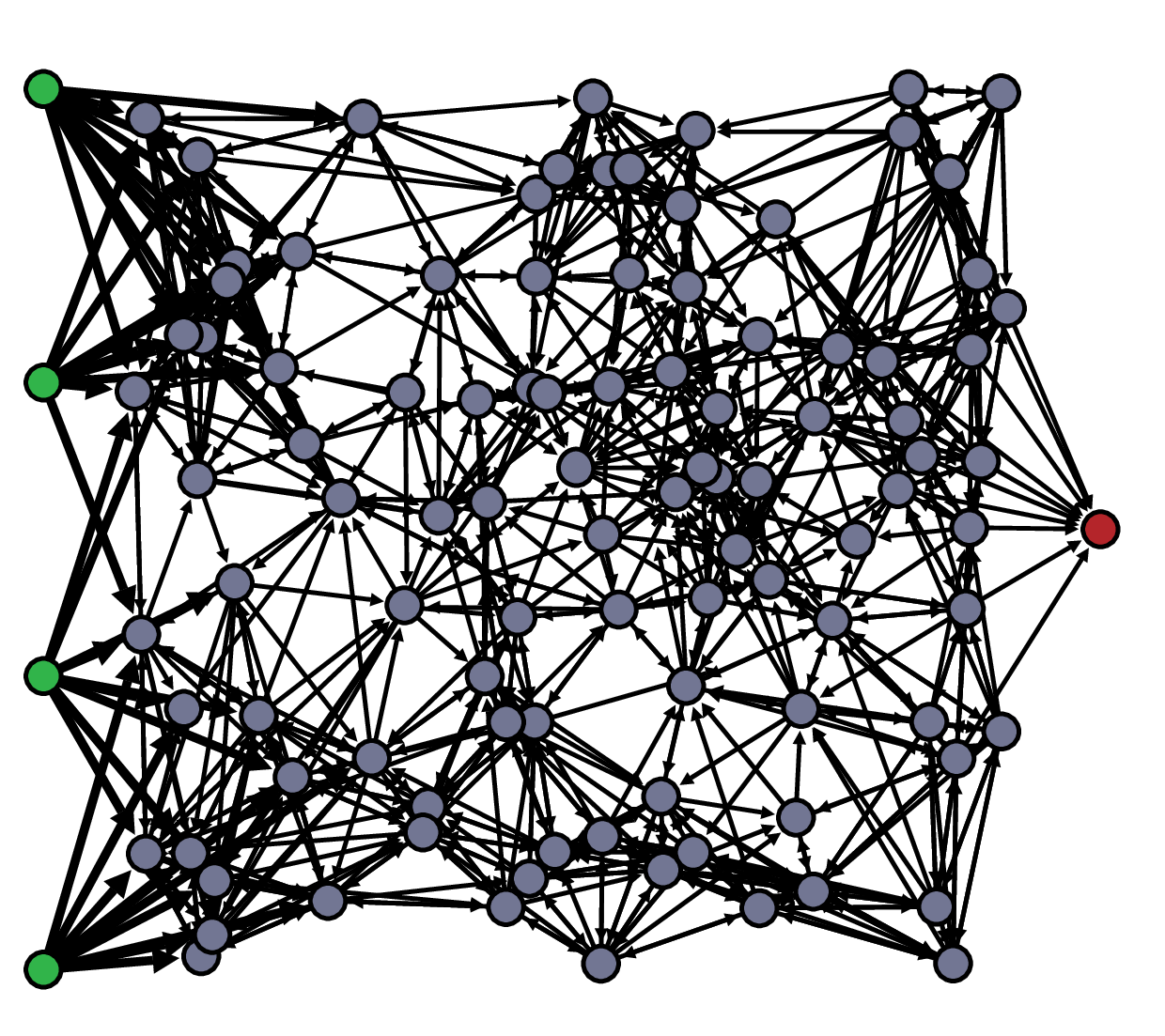}  
    \caption{ {\bf Network architecture:} Network with one hundred neurons in the network (gray neurons). The four input neurons (green) on the left are labeled 1 to 4 (top to bottom) and are connected to the network. On the right is the output neuron (red) which senses incoming signals from its ten nearest nodes. The strength of the synapses is indicated by their thickness. }
  \label{network}
\end{figure}

The network consists of 4 input neurons, 1 output neuron and $N$ neurons in the hidden network, which are placed randomly in a square, with side length $L$. The area is chosen to scale proportional to $N$, such that the density of neurons remains constant  (Fig. \ref{network}). In the further discussions we will use $N$ to characterize the size of the network. Each neuron in the hidden network has ten outward connections. Each connection is established by choosing a distance $d$ from an exponential distribution \cite{roerig2002relationships} $p(d) = \frac{1}{d_0} e^{ -d / d_0}$ and searching for a neuron at a distance sufficiently close to $d$. The input neurons also have ten outward connections with their nearest neighbours whereas the output neuron has no outward but only inward connections from its ten nearest nodes. When inhibitory neurons are considered, a fixed fraction $p_{inh}$ of the neurons will be inhibitory.

\subsection*{Neuronal model}
For the firing dynamics we choose an integrate-and-fire model with discrete time steps. During each time step all neurons with potential exceeding a certain threshold $v_i \geq v_{max} = 1.0$ fire. After firing the potential of the neuron $i$ is set back to zero and the voltage of all connected neurons $j$ becomes
\begin{align}
v_j(t+1) = v_j(t) \pm \omega_{ij} \eta_i
\label{eq1}
\end{align}
where $\omega_{ij}$ is the synaptic strength and $\eta_i \in [0,1]$ stands for the amount of releasable neurotransmitter, which for simplicity is the same for all synapses of neuron $i$. The plus or minus sign is for excitatory or inhibitory synapses, respectively. After each firing $\eta_i$ decreases by a fixed amount $\Delta \eta = 0.2$, which guarantees that the network activity will decay in a finite time. The initial conditions are $v=0$ and $\eta=1$ for all neurons. All neurons which have reached a potential higher than the threshold will fire at the next time step. After firing a neuron will be in a refractory state for $t_{refr}$ time step, during which it cannot receive or send any signal.

\subsection*{Learning}
Learning binary input-output relations such as the XOR-rule has been a classical benchmark problem for neural networks. In our case the XOR-rule turned out to be too simple as only very small networks with unfavorable parameters were not always able to learn it. We choose to study a more complex problem with up to $15$ different Boolean functions of $4$ binary inputs (see Table \ref{rules}). We will study how our learning procedure performs for different parameters and network sizes. 

\begin{table}[htbp]
\begin{center}
\begin{tabular}{ l  c c c c c c c c c c c c c c c }
\toprule
Input 1 & 1 & 0 & 1 & 0 & 0 & 0 & 1 & 1 & 1 & 1 & 0 & 0 & 1 & 1 & 0\\

Input 2 & 0 & 1 & 1 & 0 & 0 & 0 & 1 & 0 & 1 & 0 & 1 & 1 & 1 & 0 & 1\\

Input 3 & 0 & 0 & 0 & 1 & 0 & 1 & 1 & 1 & 1 & 0 & 1 & 0 & 0 & 1 & 1\\

Input 4 & 0 & 0 & 0 & 0 & 1 & 1 & 1 & 0 & 0 & 1 & 0 & 1 & 1 & 1 & 1\\
\midrule
Output & 1 & 1 & 0 & 1 & 1 & 0 & 0 & 1 & 0 & 1 & 0 & 1 & 0 & 1 & 0\\
\bottomrule
\end{tabular}
\end{center}
\caption{Input-output relations to be learned by the network.}
\label{rules}
\end{table}%

Each input bit is applied to one input neuron by making it fire or not fire if the input is one or zero, respectively. The activity then propagates through the network and the binary output value 1 is identified with the output neuron firing at least once and zero if it does not fire during the whole network activity. 

Whether the activity reaches the output neuron depends solely on the strength of the synapses. If the average strength of the synapses is too low, the propagation of the activity stops soon and only a fraction of the entire network will fire. On the other hand for very strong synapses almost all neurons will fire several times and the output neuron will fire for each input pattern. Therefore an optimal choice for the average synaptic strength is a "critical value" where on average for each input pattern there is an equal chance that the output neuron fires or does not fire. We approximate this "critical point" by initially choosing weak synapses ($\omega_{ij} = 1.0$ for the synapses of the input neurons and $\omega_{ij} = 0.1$ for all other synapses) and when applying the input patterns, as long as no input leads to firing of the output neuron, all synapses are strengthened by a small amount ($0.001$ times the actual strength of the synapse). As soon as the first input leads to the output neuron firing, the network is considered to be close to the "critical point" and the learning procedure starts.

Learning proceeds as follows: All inputs of the rule are fed into the network one after the other, where the value of the neurotransmitters and the voltages are restored back to $\eta = 1.0$ and $v = 0$ before each input. Whenever the result is wrong the synaptic strengths are adapted. If the correct result is one but the output neuron did not fire, synapses are strengthened, otherwise if the correct result is zero but the output neuron fired, synapses are weakened \cite{BakChialvo}. Only those synapses activated during the activity propagation are modified. After a neuron fires all synapses which do not lead to neurons which were in the refractory state are considered as activated. The adaptation signal is assumed to be released at the output and decrease exponentially in space
\begin{align}
\Delta \omega_{ij} =  \pm \alpha \omega_{ij}  n_{act} e^{-r/r_0}
\end{align}
where $\alpha=0.001$ is the adaptation strength and $r$ is the euclidean distance between the output neuron and the postsynaptic neuron of the synapse, neuron $j$. The postsynaptic neuron is chosen based on the assumption that a synapse is located much closer to the postsynaptic than to the presynaptic neuron. Different values for $\alpha$ have been tested and are found to affect the learning speed but not the qualitatively results. The plus or minus sign is for excitatory or inhibitory synapses, respectively. Additionally the adaptation is proportional to the strength of the synapse itself and to the number of times the synapse was activated ($n_{act}$). The $\omega_{ij}$ and the $n_{act}$ dependence are considered since they lead to an increase in learning performance. A maximum synaptic strength ($\omega_{max}=2$) is imposed. 
If the activity is so weak that the voltage of the output neuron remains unchanged, the system is not able to provide an answer and therefore we increase all the weights by a small amount ($\Delta \omega_{ij} =  \alpha \omega_{ij}$).

\subsection*{XOR gate with only excitatory neurons}
\begin{figure}[ht]
  \centering
    \includegraphics[width=0.6\textwidth]{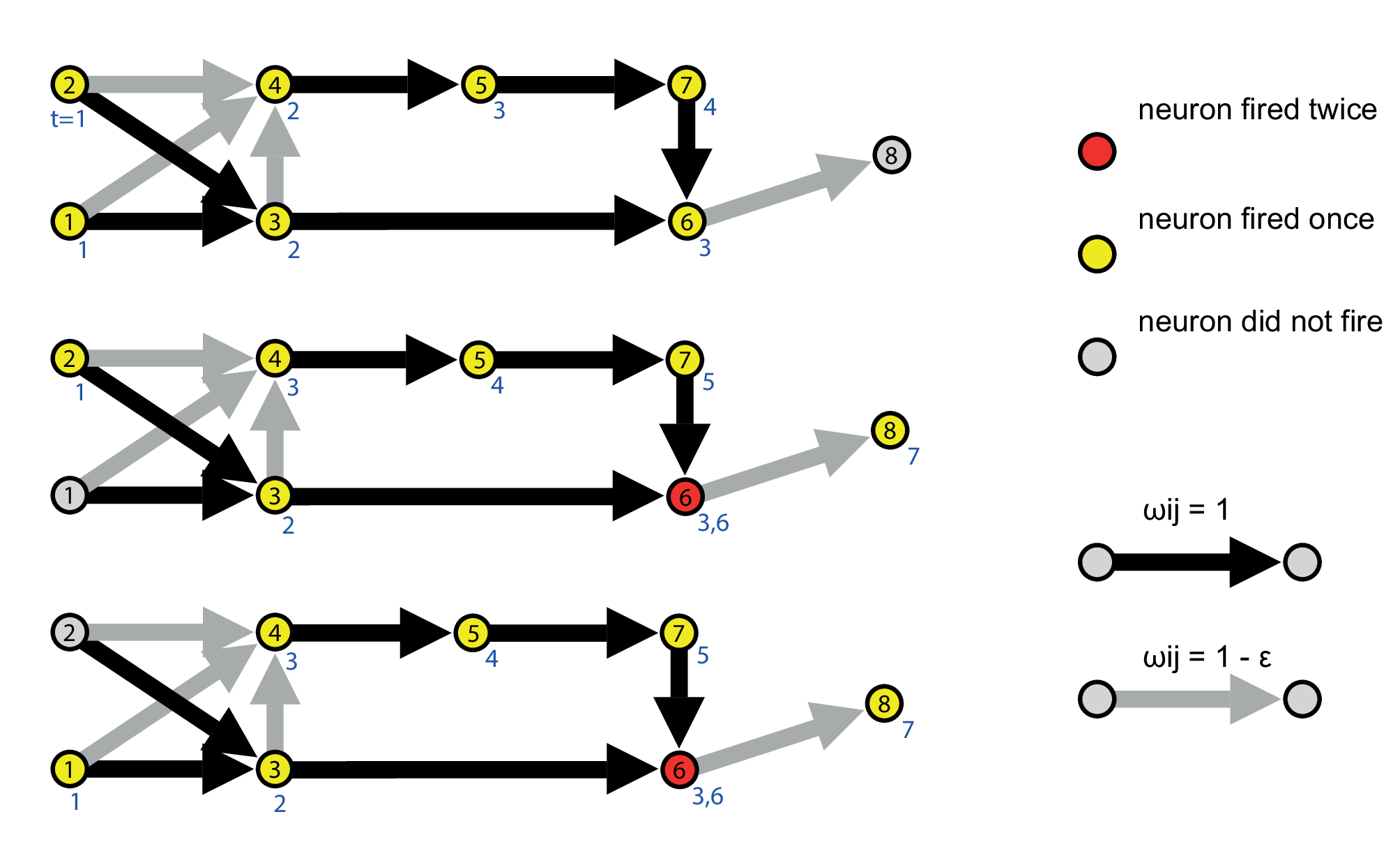}  
    \caption{ {\bf XOR gate with only excitatory neurons.}  The figure above shows a possibility for a XOR gate with only excitatory neurons with a refractory period of one time-step. The blue numbers below each neuron indicate at which time step the neuron fires.}
  \label{xor}
\end{figure}

A biological neural network consists of excitatory ($\omega_{ij} \ge 0$) and inhibitory ($\omega_{ij} \le 0$) synapses. It has actually been shown that a certain percentage of inhibitory synapses can increase the learning performance \cite{LucillaOptimal}. The importance of inhibiting signals can be seen by looking at the XOR-rule (Fig. \ref{xor}): When one of the two input neurons fires the output neuron should fire as well, but when both input neurons fire the output neuron should not fire. Using only synapses with positive weights we might intuitively think that by the principle of superposition a stronger input should always lead to a stronger output. This, however, does not hold in general. An interesting and important ingredient to generate inhibition is the refractory time. Figure \ref{xor} shows an example of a small network that performs the XOR rule with excitatory neurons and a refractory time of one time step. Black and gray arrows are for weights equal to and barely less than one, respectively. When both input neurons (neuron 1 and 2 in the figure) are stimulated neuron 4 fires one time step earlier than in the other two cases, where neuron 4 fires after a second stimulation by neuron 3. This earlier firing of neuron 4 leads to neuron 7 stimulating neuron 6 when it is in the refractory period. Therefore, neuron 6 only fires once which is not sufficient to make the output neuron (neuron 8) fire. The self-organization of the learning mechanism can take advantage of this mechanism to suppress certain signals. Therefore, even if  purely excitatory network is not biologically reasonable, the refractory time can induce inhibition which can be of computational relevance. Therefore, we chose to perform most simulations with purely excitatory neurons.

\section*{Results}
Our learning procedure runs until either sequentially all input-output relations are learned or a maximal number of learning steps $T_{max}$ has been performed, where a learning step occurs whenever the network generates a wrong result for the actual input. For most of the simulations the networks will be trained to learn the first ten patterns from Table 1. The learning performance is evaluated by generating a statistical ensemble of $2000$ random networks with the same parameters and the success rate $s$ is defined by the ratio of networks which are able to learn all patterns within $T_{max}$.

\begin{figure}[!ht]
    \includegraphics[width=1\textwidth]{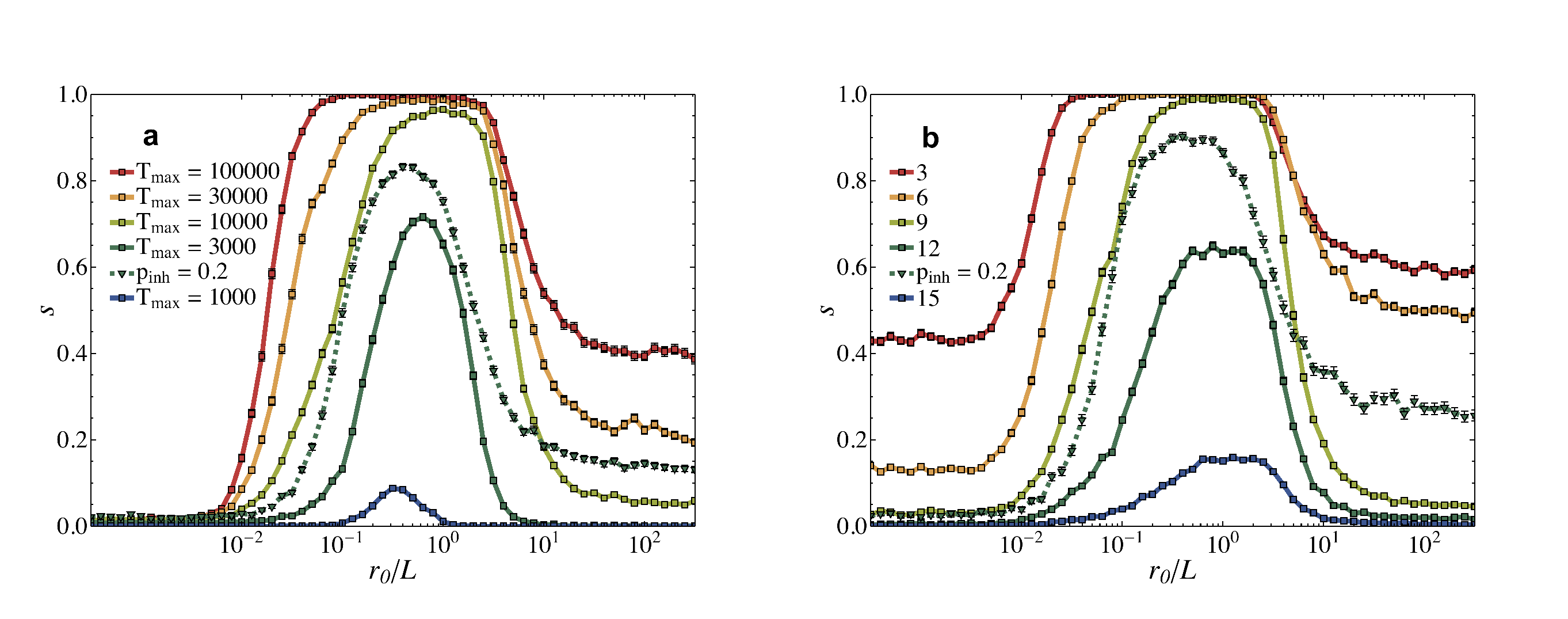}  
 \caption{ {\bf Ranges of the learning length optimizing the learning performance.} The success rate $s$ as a function of $r_0/L$. ({\bf a}) Different maximal learning times $T_{max}$ show a consistent improvement in performance ($N=1000$, $d_0 = 2$,$t_{refr}=1$, first ten patterns of Table 1). ({\bf b}) The performance dependence on the number of patterns the system has to learn ($T_{max} = 10000$, $N=1000$,$t_{refr}=1$, $d_0 = 2$). For $T_{max}=3000$ (in Fig. {\bf a}) and for 12 patterns (in Fig. {\bf b}) we also performed simulations with inhibitory neurons $p_{inh}=0.2$ resulting in an overall performance improvement. The error bar represent a $95 \%$ confidence interval.}
  \label{learningTime}
\end{figure}

We first investigate the role of the learning length $r_0$ on the success rate $s$ (see Fig. \ref{learningTime}a). For very localized learning signals ($r_0 < 10^{-1}$) the success rates are close to zero. This is not surprising as only the synapses leading to the output neuron are modified by a noticeable amount. For larger $r_0$ the learning adaptation penetrates deeper into the network and the success rate rapidly increases up to $s = 1.0 $ for $r_0 = 10$ and $T_{max} = 100,000$. Interestingly, when $r_0$ exceeds the system size ($r_0/L = 1$), which implies that all synapses in the network are adapted, the learning performance strongly decreases. Overall, we find that the learning performance can strongly be optimized by the spacial extent of the teaching signal.

Figure \ref{learningTime}b shows the learning performance for a different number of patterns to be learned by the network. Unsurprisingly the performance is higher when only the first three patterns from Table 1 have to be learned compared to the case when all the fifteen patterns have to be learned. It is interesting to notice that the value for $r_0$ at which the learning performance increases shifts towards larger values the more patterns the network has to learn. This suggests that the more patterns a network has to learn the deeper the learning signal has to penetrate into the network.

\begin{figure}[!ht]
    \includegraphics[width=1\textwidth]{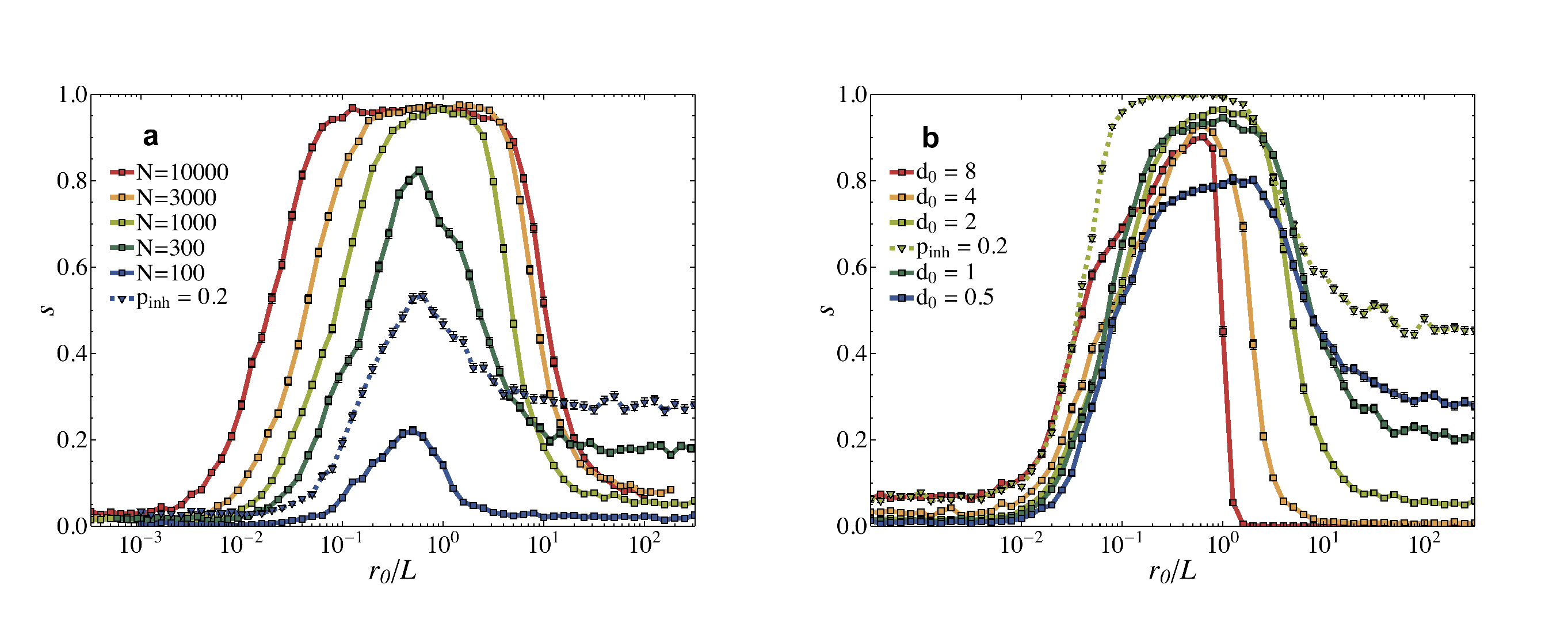}  
    \caption{ {\bf System size dependence and influence of the synaptic length.} ({\bf a}) The success rate $s$ as a function of $r_0/L$ for different network sizes $N$ ($T_{max} = 10000$, $d_0 = 2$,$t_{refr}=1$, first ten patterns of Table 1). Larger networks show a better performance and the peak of the performance roughly moves proportionally to $L$. ({\bf b}) The effect of the network structure is analyzed by plotting $s$ for different synaptic characteristic length $d_0$ ($N=1000$, $T_{max} = 10000$,$t_{refr}=1$, first ten patterns of Table 1). For $N=100$ (in Fig. {\bf a}) and for $d_0 = 2$ (in Fig. {\bf b}) we also performed simulations with inhibitory neurons $p_{inh}=0.2$ resulting in an overall performance improvement. The error bars represent a $95 \%$ confidence interval.}
  \label{networkSize}
\end{figure}

Next we investigate, the role of the network size on the performance. Indeed, we find a monotonic increase in performance (see Fig. \ref{networkSize}a). Interestingly the performance exhibits its maximum at a value of $r_0$ increasing proportionally to $L$, whereas the s-decay shifts towards larger $r_0$ for larger system size. This result confirms that the learning capabilities decrease when $r_0$ exceeds the system size $L$ and the adaptation changes all activated synapses in an almost uniform way.  In Fig. \ref{networkSize}b we vary the characteristic length for synaptic connections $d_0$ and study its effect on the learning performance. When $r_0$ exceeds the system length, $d_0$ strongly affects the learning performance. For longer synapses (higher $d_0$) the performance drops much faster as $r_0$ increases. This result is in agreement with the abrupt decrease of the performance for plastic adaptation acting over a range larger than the system size $L$. For very large $d_0$, the synaptic network also undergoes quite uniform plastic adaptation over large distances. The maximum performance appears to weakly depend on $d_0$, except for very localized connections.  

We also ask the question whether the space dependence of the adaptation influences the presented results. Therefore we perform simulations using a plastic adaptation that decrease in space as a Gaussian. Results are very similar (data not shown) to those obtained for a synaptic adaptation decaying exponentially, except that for the Gaussian adaptation the peaks are slightly narrower.

\begin{figure}[!ht]
    \includegraphics[width=1\textwidth]{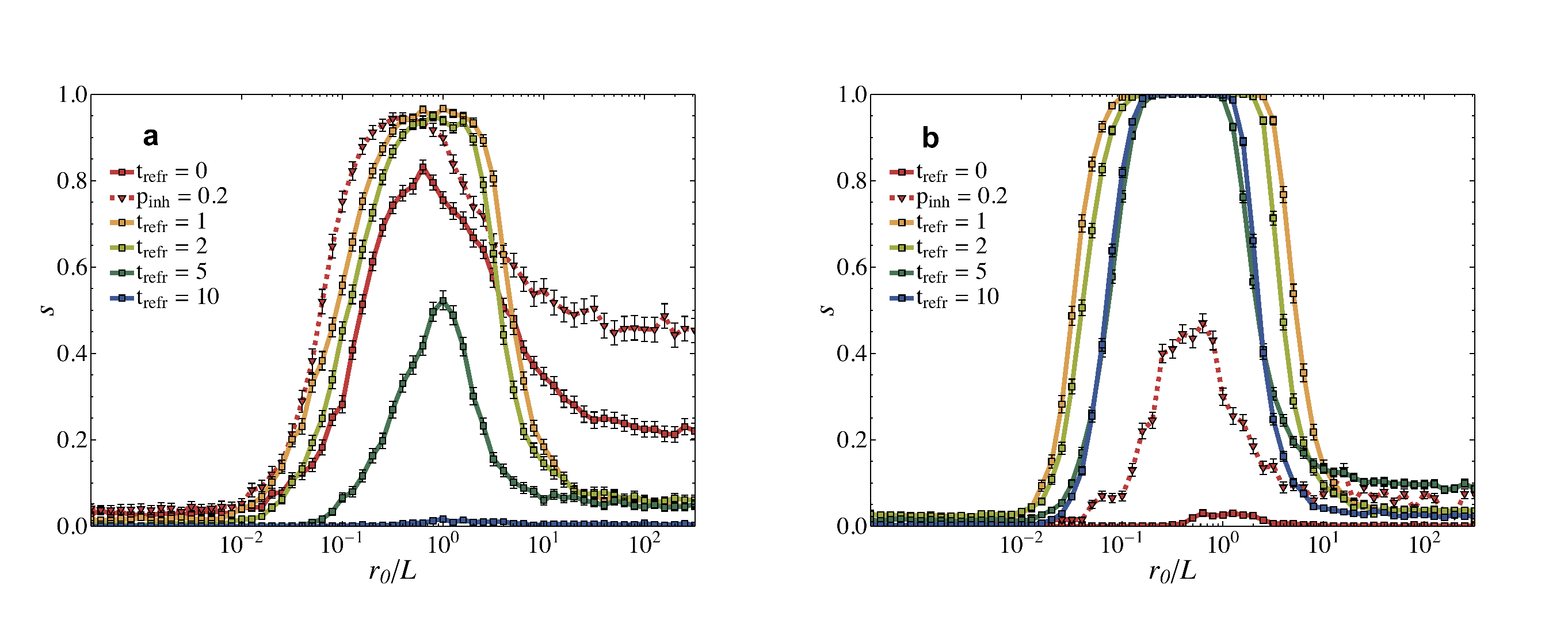}  
    \caption{ {\bf Role of the refractory time on the performance.} ({\bf a}) The success rate $s$ for different refractory times, ranging from $t_{refr} = 0$ to  $t_{refr} = 10$ ($N=1000$, $T_{max} = 10000$, $d_0 = 2$, first ten patterns out of Table 1). ({\bf b}) Performance obtained for the neuronal activation function given in Eq. \ref{newActivation}. For $t_{refr}=0$ (in Fig. {\bf a} and {\bf b}) we also performed simulations with inhibitory neurons $p_{inh}=0.2$ resulting in an overall performance improvement. The error bars represent a $95 \%$ confidence interval.}
  \label{refractoryTime}
\end{figure}

Further, we tested the influence of the refractory time on the learning performance (see Fig. \ref{refractoryTime}a). When the refractory time is zero learning is possible, but the performance is worse compared to $t_{refr} = 1$ and $t_{refr} = 2$. For $t_{refr} = 10$ the performance decreases almost to zero. This might be related to the fact that for larger refractory times it becomes less likely that a single neuron fires more than once, therefore preventing the occurrence of activity loops. This result shows that the refractory time is an important ingredient to generate inhibition, but also the activation function (Heaviside step function) causes dissipation which can create inhibition. We could reduce this type of dissipation by using an activation function that is proportional to the voltage of the firing neuron. Equation \ref{eq1} changes then to:
\begin{align}
v_j(t+1) = v_j(t) \pm \omega_{ij} \eta_i v_i
\label{newActivation}
\end{align}
To eliminate dissipation completely we would need to set the firing threshold to zero. However, this would make learning impossible as every input will make all successive neurons fire and nothing can stop the output neuron from firing. In Fig. \ref{refractoryTime}b we show the effect of the linear activation function (eq. \ref{newActivation}) on the learning performance. The result confirms our assumption. For a refractory time equal to zero the learning performance drops to zero. 
For refractory times larger than zero the overall performance is higher compared to the Heaviside activation function. The performance also does not decrease for larger refractory times. Adding inhibitory neurons makes it possible for neurons with $t_{refr}=0$ to learn, but with a low performance.

In each case of Fig. \ref{learningTime} - \ref{refractoryTime} we perform the same simulation, but with a fraction $p_{inh} = 0.2$ of inhibitory neurons. Inhibitory neurons noticeably increase the maximum learning performance and the performance decay for larger $r_0$ is slightly less pronounced. In Fig. \ref{refractoryTime}b the inhibitory neurons make it possible to learn if the refractory time is zero, which is not possible without inhibitory neurons.  

\section*{Conclusions}
We show that a neural network trained by a distance dependent plastic adaptation can learn Boolean rules with a very good performance. The spatial extent of the learning signal was found to have a huge impact on the learning performance. A very localized plastic adaptation, which only modifies the synapses directly connected to the output neuron, results, unsurprisingly, in poor performance. For deeper adaptation signal, the performance strongly increases, until the learning length $r_0$ exceeds the system length $L$, where the performance decreases. When $r_0 \gg L$ the adaptation signal becomes quite uniform in space and the lack of variability might limit the learning capabilities. Similar behaviour is observed for increasing characteristic length of synaptic connections. Indeed, current evidence from functional magnetic resonance imaging (fMRI) experiments \cite{garrett2010blood,garrett2011importance} and EEG data \cite{ghosh2008noise,mcintosh2008increased} shows that a greater brain signal variability indicates a more sophisticated neural system, able to
explore multiple functional states. Signal variability also reflects a
greater coherence between regions and a natural balance between
excitation and inhibition originating the inherently variable response
in neural functions. Furthermore, the observation that older adults
exhibit less variability reflecting less network complexity and integration,
suggests that variability can be a proxy for optimal systems.

Interestingly, we show that a network with only excitatory neurons can learn non-linearly separable problems with a success rate of up to $100 \%$. In this case, the neural refractory time generates inhibition in the neuronal model. If we include additional inhibitory neurons the performance still increase noticeably.

Regarding the network structure we find that the synaptic length also has an impact on the performance: For networks dominated by local connections the performance is worse compared to networks with long range connections. We conclude that the structure of the network and the locality of the plastic adaptation have an important role in the learning performance. To assess the generality of our results it will be necessary to study more complex forms of learning like, for example, pattern recognition. Further investigations should also try to improve the learning performance to become comparable to machine learning algorithms.


\section*{Acknowledgements}
The authors thank L.M. van Kessenich for many interesting discussions. We acknowledge financial support from the European Research Council (ERC) Advanced Grant FP7-319968 FlowCCS.

\section*{Author contributions statement}

D.L.B conceived the research and conducted the numerical simulations. L.d.A and H.J.H. gave important advise. All authors contributed to the writing of the manuscript.

\section*{Additional information}

\textbf{Competing financial interests} The authors declare no competing financial interests.

\end{document}